\documentclass[11pt]{article}
\usepackage[dvips]{graphicx}  
\begin{document}
\title{Black Holes in Elliptical and Spiral Galaxies and in 
Globular Clusters}

\begin{center}

{\Large\bf Black Holes in Elliptical and Spiral Galaxies and in 
Globular Clusters
\rule{0pt}{13pt}}\par
\bigskip
Reginald T. Cahill \\ 
{\small\it School of Chemistry, Physics and Earth Sciences, Flinders University,\\
Adelaide 5001, Australia\rule{0pt}{13pt}}\\
\raisebox{-1pt}{\footnotesize E-mail: Reg.Cahill@flinders.edu.au}\\physics/0508175
\\Published: Progress in Physics, {\bf 3}, 51-56(2005)\par
\bigskip\smallskip
{\small\parbox{11cm}{%
Supermassive black holes have been discovered at the centers of  galaxies, and also in globular clusters.
The data shows correlations between the black hole mass and the elliptical galaxy mass  or globular cluster mass.  
It is shown that
this correlation is accurately predicted by a theory of gravity which includes the new dynamics of self-interacting  space.
In spiral galaxies this dynamics is  shown to explain the so-called `dark matter'  rotation-curve anomaly, 
and also explains the earth based bore-hole $g$ anomaly data. Together these effects imply 
that the strength of the self-interaction dynamics is determined by the fine structure constant. 
This has major implications for fundamental physics and cosmology.   \rule[0pt]{0pt}{0pt}}}\bigskip
\end{center}


\section{Introduction}
\begin{figure*}[t]
\hspace{5mm}\includegraphics[scale=0.44]{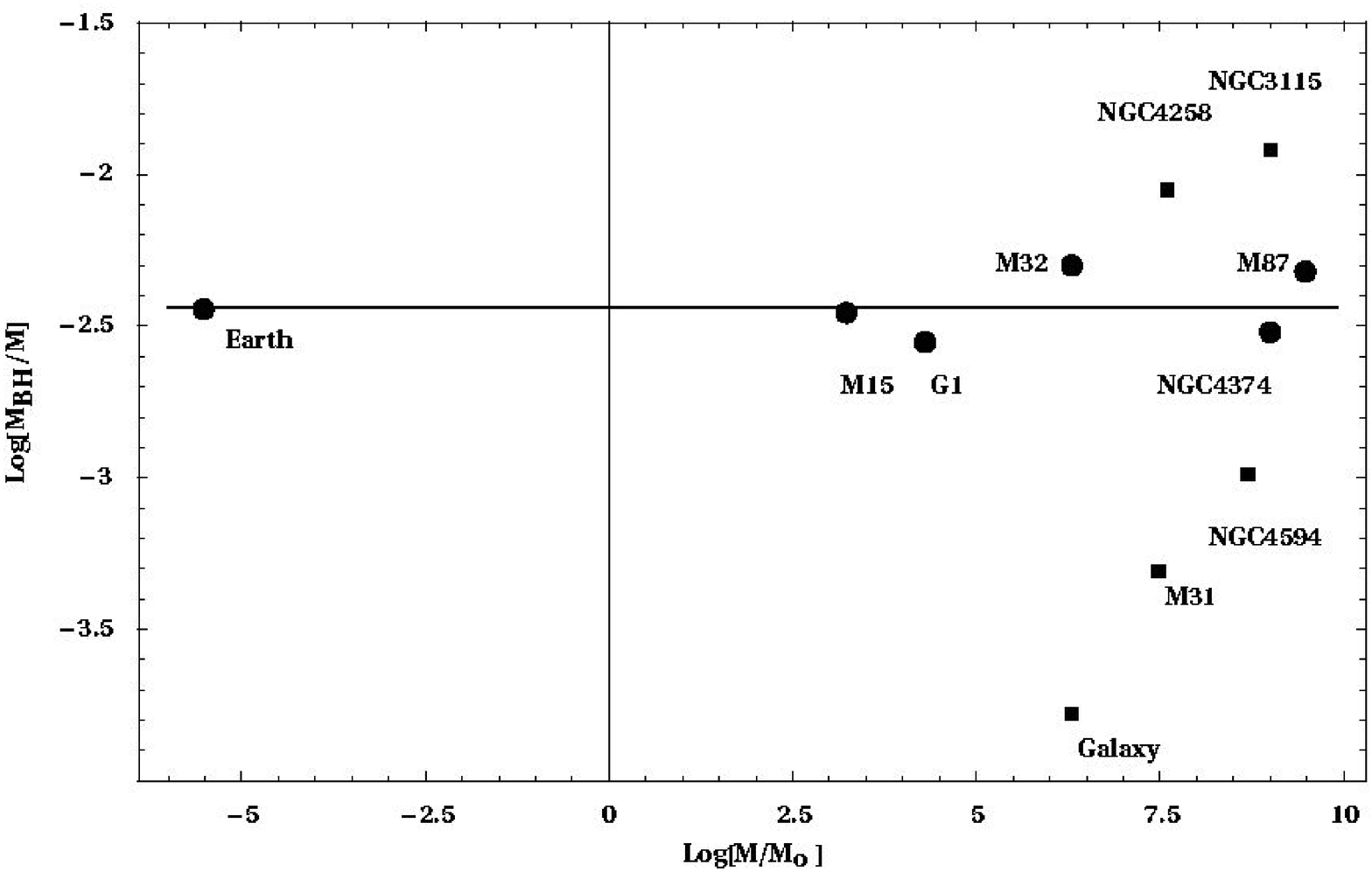}
\caption{\small{The data shows $\mbox{Log}_{10}[M_{BH}/M]$ for the `black hole' or `dark matter' masses $M_{BH}$  for a
variety of spherical matter systems with masses $M$, shown by solid circles, plotted against 
$\mbox{Log}_{10}[M/M_0]$, where
$M_0$ is the solar mass, showing agreement with the `$\alpha/2$-line' ($\mbox{Log}_{10}[\alpha/2]=-2.44$) predicted by
(\ref{eqn:TotalDM}), and ranging over 15 orders of magnitude. The `black hole' effect is the same phenomenon as the
`dark matter' effect. The data ranges from the earth, as observed by the bore hole $g$ anomaly, to globular cluster
M15 \cite{GlobularM15, M15Mass} and G1 \cite{GlobularG1}, and then to spherical `elliptical' galaxies M32 (E2), NGC
4374 (E1)  and M87 (E0). Best fit to the data from these star systems gives $\alpha=1/134$, while for the earth data
in Fig.\ref{fig:Greenland}
$\alpha=1/139$. A best fit to all the spherical  systems in the plot gives $\alpha=1/136$. In these systems the `dark
matter' or `black hole' spatial self-interaction effect is induced by the matter. For the spiral galaxies, shown by
the filled boxes, where here $M$ is  the bulge mass, the black hole masses do not correlate with the
`$\alpha/2$-line'. This is because these systems form by matter in-falling to a primordial black hole, and so these
systems are more contingent. For spiral galaxies this dynamical effect manifests most clearly via the non-Keplerian
rotation-velocity curve, which decrease asymptotically very slowly, as shown in Fig.\ref{fig:NGC3198}, as determined
by the small value of
$\alpha\approx 1/137$. The galaxy data is from Table 1 of \cite[updated]{Kormendy}.}
\label{fig:blackholes}}\end{figure*}

Our understanding of gravity is based on Newton's modelling of Kepler's phenomenological laws 
for the motion of the planets within the solar system.  In this model Newton took the gravitational acceleration field
to be the fundamental dynamical degree of freedom, and which is determined by the matter distribution; essentially via
the `universal inverse square law'. However  the observed linear correlation between masses of
black holes  with the masses of the `host' elliptical galaxies or globular clusters  suggests that
either the formation of these systems involves common evolutionary dynamical processes or that perhaps some new aspect
to gravity is being revealed.   Here it is shown that if rather than an acceleration field a velocity field is assumed
to be fundamental to gravity, then we immediately find that these black hole effects arise as a space
 self-interaction dynamical effect, and that the observed correlation is simply that $M_{BH}/M= \alpha/2$
 for spherical systems, where
$\alpha$ is the fine structure constant $(
\alpha=e^2/\hbar c=1/137.036)$, as shown in Fig.\ref{fig:blackholes}.  This dynamics also
manifests within the earth, as revealed by the bore hole $g$ anomaly data, as in Fig.\ref{fig:Greenland}. It  also 
offers an explanation  of the `dark matter' rotation-velocity effect, as illustrated in Fig.\ref{fig:NGC3198}. 
This common explanation for a range of seemingly unrelated effects has deep implications for fundamental physics and
cosmology.

\section{Modelling Gravity}

Let us phenomenologically investigate the consequences of using a velocity field ${\bf v}({\bf r},t)$  to be the
fundamental dynamical degree of freedom to model gravity.  The  gravitational acceleration field is then defined  by
the Euler form
\begin{eqnarray}
{\bf g}({\bf r},t)&\equiv&\lim_{\Delta t \rightarrow 0}\frac{{\bf v}({\bf r}+{\bf v}({\bf r},t)\Delta t,t+\Delta
t)-{\bf v}({\bf r},t)}{\Delta t} \nonumber \\
&=&\frac{\partial {\bf v}}{\partial t}+({\bf v}.\nabla ){\bf v}
\label{eqn:basicg}\end{eqnarray} 
This form is mandated by Galilean covariance under change of observer.  
A minimalist non-relativistic modelling of the dynamics for this velocity field gives a direct account of the various
phenomena noted above; basically the Newtonian formulation of gravity missed a key dynamical effect that did not
manifest within the solar system.  

In terms of the velocity field Newtonian gravity dynamics involves using $\nabla.$ to construct a rank-0 tensor that
can be related to the matter density
$\rho$. The coefficient turns out to be the Newtonian gravitational constant
$G$.
\begin{equation}
\nabla.\left(\frac{\partial {\bf v} }{\partial t}+({\bf
v}.{\bf \nabla}){\bf v}\right)=-4\pi G\rho,
\label{eqn:CG1}\end{equation}
This is clearly  equivalent to the differential form of Newtonian gravity,
$\nabla.{\bf g}=-4\pi G \rho$. Outside of a spherical mass $M$ (\ref{eqn:CG1}) has solution (we assume
$\nabla \times {\bf v}={\bf 0}$, then $({\bf v}.\nabla){\bf v}=\frac{1}{2}\nabla({\bf v}^2)$)
\begin{equation}
{\bf v}({\bf r})=-\sqrt{\frac{2GM}{r}}\hat{\bf r},
\label{eqn:vin}\end{equation} 
for which  (\ref{eqn:basicg}) gives the usual inverse  square law
\begin{equation}
{\bf g}({\bf r})=-\frac{GM}{r^2}\hat{\bf r}.
\label{eqn:InverseSqLaw}\end{equation} 
The  simplest non-Newtonian dynamics involves the two rank-0 tensors constructed at 2nd order from $\partial
v_i/\partial x_j$
\begin{equation}
\nabla.\left(\frac{\partial {\bf v} }{\partial t}+({\bf v}.{\bf \nabla}){\bf v}\right)
+\frac{\alpha}{8}(tr D)^2+\frac{\beta}{8}tr(D^2)=-4\pi G\rho,
\label{eqn:CG}\end{equation}
\begin{equation} D_{ij}=\frac{1}{2}\left(\frac{\partial v_i}{\partial x_j}+\frac{\partial v_j}{\partial x_i}\right),
\label{eqn:Ddefn1}\end{equation}
and involves two arbitrary dimensionless constants. The velocity  in (\ref{eqn:vin}) is also a solution to
(\ref{eqn:CG})  if $\beta=-\alpha$, and we then define 
\begin{equation}
C({\bf v},t)=\frac{\alpha}{8}((tr D)^2-tr(D^2)).
\label{eqn:Cterm}\end{equation}
Hence the modelling of gravity by (\ref{eqn:CG}) and (\ref{eqn:basicg}) now involves two gravitational constants $G$
and
$\alpha$, with $\alpha$ being the strength of the   self-interaction dynamics, but which was not apparent  in the solar
system dynamics. We now show that all the various phenomena discussed herein imply that $\alpha$ is the fine
structure constant $\approx 1/137$ up to experimental errors \cite{Cahill}. Hence non-relativistic gravity is a more complex
phenomenon than currently understood. The new key feature is that (\ref{eqn:CG}) has a one-parameter $\mu$ class of vacuum
($\rho=0$) `black hole' solutions in which the velocity field self-consistently maintains the singular form
\begin{equation}
{\bf v}({\bf r})=-\mu r^{-\alpha/4}\hat{\bf r}.
\label{eqn:blackholeflow}\end{equation}  
This class of solutions will be seen to account for the `black holes' observed in galaxies and globular cluster. As
well this velocity field,  from (\ref{eqn:basicg}), gives rise to a non- `inverse square law' acceleration
\begin{equation}
{\bf g}({\bf r})=-\frac{\alpha\mu}{4} r^{-(1+\alpha/4)}\hat{\bf r}.
\label{eqn:dmg}\end{equation}
This turns out to be the cause of the so-called `dark-matter' effect observed in spiral galaxies.  For this reason we
define
\begin{equation}
\rho_{DM}({\bf r})=\frac{\alpha}{32\pi G}( (tr D)^2-tr(D^2)),  
\label{eqn:DMdensity}\end{equation}
 so that  (\ref{eqn:CG}) and (\ref{eqn:basicg}) can be written as
\begin{equation}\label{eqn:g2}
\nabla.{\bf g}=-4\pi G\rho-4\pi G \rho_{DM},
\end{equation}
which shows that we can think of the new self-interaction dynamics as generating an effective `dark matter' density.

\section{Spherical Systems} 
It is sufficient here to consider time-independent and spherically symmetric solutions of  (\ref{eqn:CG}) for which
$v$ is radial.
 Then we have the integro-differential form for (\ref{eqn:CG})
\begin{equation}\label{eqn:integraleqn}
v^2(r)=
2G\int d^3
s\frac{\rho(s)+\rho_{DM}(  v(s))}{|{\bf r}-{\bf s}|},
\end{equation} 
\begin{equation}
\rho_{DM}(v(r))= \frac{\alpha}{8\pi G}\left(\frac{v^2}{2r^2}+ \frac{vv^\prime}{r}\right).
\label{eqn:dm1}\end{equation}
as $\nabla^2 \frac{1}{|{\bf r}-{\bf s}|}=-4\pi\delta^4({\bf r}-{\bf s}) $. This then gives
\begin{eqnarray}
v^2(r)&=&\frac{8\pi G}{r}\int_0^r  s^2ds \left[\rho(s)+\rho_{DM}(v(s))\right]+
\nonumber \\&& 8\pi G\int_r^\infty sds
\left[\rho(s)+\rho_{DM}(v(s))\right]
\label{eqn:integralEqn}\end{eqnarray}
on doing the angle integrations.
We can also write  (\ref{eqn:CG}) as a non-linear differential equation
\begin{equation}
2\frac{vv^\prime}{r} +(v^\prime)^2 + vv^{\prime\prime} =-4\pi G\rho(r)-4\pi G \rho_{DM}(v(r)). 
\label{eqn:InFlowRadial}
\end{equation}

\section{Minimal Black Hole Systems}
 There are two classes of solutions when matter is present. The simplest is when the black hole forms as a consequence
of the velocity field generated by the matter, this generates what can be termed an induced minimal black hole. This is
in the main applicable to systems such as planets, stars, globular clusters and elliptical  galaxies.  The second class
of solutions correspond to non-minimal black hole systems; these arise when the matter congregates around a
pre-existing `vacuum' black hole.  The minimal black holes are simpler to deal with, particularly when the matter
system is spherically symmetric.  In this case the non-Newtonian gravitational effects are confined to within the
system. A simple way to arrive at this property is to solve (\ref{eqn:integralEqn}) perturbatively. When the
matter density is confined to a sphere of radius $R$ we find on iterating (\ref{eqn:integralEqn}) that the `dark
matter' density is confined to that sphere, and that consequently $g(r)$ has an inverse square law behaviour
outside of the sphere. Iterating (\ref{eqn:integralEqn}) once  we find inside radius $R$ that 
\begin{equation}
\rho_{DM}(r)=\frac{\alpha}{2r^2}\int_r^R s\rho(s)ds+O(\alpha^2).
\label{eqn:perturbative}\end{equation}
 and that  the total `dark matter'
\begin{eqnarray}
M_{DM}&\equiv& 4\pi\int_0^R r^2dr\rho_{DM}(r)\nonumber \\&=&\frac{4\pi\alpha}{2}\int_0^R
r^2dr\rho(r)+O(\alpha^2)\nonumber \\
&=&\frac{\alpha}{2}M+O(\alpha^2),
\label{eqn:TotalDM}\end{eqnarray}
where $M$ is the total amount of (actual) matter. Hence to $O(\alpha)$   $M_{DM}/M=\alpha/2$ independently of the matter
density profile. This turns out to be  a very useful property as knowledge of the density profile is then not
required in order to analyse observational data. Fig.\ref{fig:blackholes} shows the value of $M_{BH}/M$ for, in
particular, globular clusters $M15$ and $G1$ and highly spherical `elliptical' galaxies M32, M87 and NGC 4374, showing
that this ratio lies close to the `$\alpha/2$-line', where $\alpha$ is the fine structure constant $\approx 1/137$.
However for the spiral galaxies their  $M_{DM}/M$  values do not cluster  close to the $\alpha/2$-line. Hence it is
suggested that these spherical systems manifest the minimal black hole dynamics outlined above.  However this dynamics
is universal, so that any spherical system must induce such a minimal black hole mode, but for which outside of such a
system only the Newtonian inverse square law would be apparent. So this mode must also apply to the earth, which is
certainly a surprising prediction. However just such an effect has manifested in measurements of $g$ in mine shafts
and bore holes since the 1980's. It will now be shown that data from these geophysical measurements give us a very
accurate determination of the value of $\alpha$ in (\ref{eqn:CG}).

\section{\bf  Bore Hole $g$ Anomaly}

\begin{figure}[t]
\hspace{40mm}\includegraphics[scale=0.9]{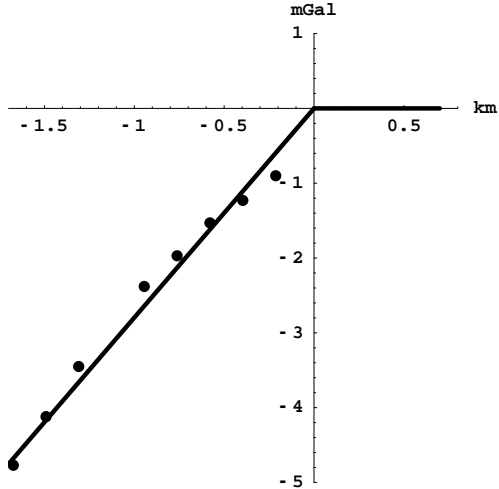}
\caption{\small{  The data shows the gravity residuals for the Greenland Ice Cap \cite{Greenland} measurements of
the
$g(r)$  profile,  defined as
$\Delta g(r) = g_{Newton}-g_{observed}$, and measured in mGal (1mGal $ =10^{-3}$ cm/s$^2$), plotted against depth in km.  Using
(\ref{eqn:deltag2}) we obtain $\alpha^{-1}=139 \pm  5 $ from fitting the slope of the data, as shown.}
\label{fig:Greenland}}\end{figure}

To understand this bore hole anomaly we need to compute the expression for $g(r)$ just beneath and just above the
surface of the earth.  To lowest order in $\alpha$ the `dark-matter' density in (\ref{eqn:perturbative}) is
substituted into  (\ref{eqn:integralEqn}) finally gives via (\ref{eqn:basicg}) the acceleration  
\begin{equation}
g(r)=\left\{ \begin{tabular}{ l} 
$\displaystyle{\frac{(1+\displaystyle{\frac{\alpha}{2}}) GM}{r^2}, \mbox{\ \ } r > R,}$  \\  
$\displaystyle{\frac{4\pi G}{r^2}\int_0^rs^2ds\rho(s) }+$ \\$\displaystyle{
\mbox{\ \ \ \ \ \ \ }\frac{2\pi\alpha G}{r^2}\int_0^r\left(\int_s^R s^\prime ds^\prime\rho(s^\prime) \right)
ds}$,\\$\displaystyle{
\mbox{\ \ \ \ \ \ \ \ \ \ \ \ \ \ \ \ \ \ \ \ \ \ \ \ \ \ \ \  } r
< R}$,\\ 
\end{tabular}\right.   
\label{eqn:ISL2}\end{equation}
This gives   Newton's `inverse square law' for $r > R$, but in which we see that the effective Newtonian gravitational
constant
 is $G_N=(1+\frac{\alpha}{2})G$, which is  different to the fundamental gravitational constant
$G$ in (\ref{eqn:CG1}). This caused by the additional `dark matter mass' in (\ref{eqn:TotalDM}). Inside the earth we
see that   (\ref{eqn:ISL2}) gives a
$g(r)$ different from Newtonian gravity. This has actually been observed in mine/borehole
measurements of
$g(r)$ \cite{Stacey,Holding,Greenland}, though of course there had been no explanation for the effect, and indeed the
reality of the effect was eventually doubted. The effect is that $g$ decreases more slowly with depth than
predicted by Newtonian gravity. Here the corresponding Newtonian form for
$g(r)$ is
\begin{equation}
g(r)_{Newton}=\left\{ \begin{tabular}{ l} 
$\displaystyle{\frac{G_N M}{r^2},\mbox{\ \ } r > R,}$  \\  
$\displaystyle{\frac{4\pi G_N}{r^2}\int_0^rs^2ds\rho(s)}$,$\displaystyle{\mbox{\ \ } r < R,}$ 
\end{tabular}\right.   
\label{eqn:earthg2}\end{equation}
with $G_N=(1+\frac{\alpha}{2})G$.  
The gravity residual  is defined as the difference between the Newtonian $g(r)$ and the measured $g(r)$, which we here
identify with the $g(r)$ from (\ref{eqn:ISL2}),
\begin{equation}
\Delta g(r)\equiv  g(r)_{Newton}-g(r)_{observed}.
\label{eqn:deltag1}\end{equation}
 Then $\Delta g(r)$ is found to be, to 1st order in $R-r$,  i.e.
near the surface, 
\begin{equation}
\Delta g(r)=\left\{ \begin{tabular}{ l} 
$\displaystyle{\mbox{\ \ }0, \mbox{\ \ } r> R,}$  \\   
$\displaystyle{-2\pi\alpha G_N\rho(R)(R-r),\mbox{\ \ } r < R,} $\\ 
\end{tabular}\right.    
\label{eqn:deltag2}\end{equation}
which is the form actually observed  \cite{Greenland}, as shown in Fig.\ref{fig:Greenland}. 

Gravity  residuals  from a bore hole
into the Greenland Ice Cap  were determined  down to a depth of 1.5km. The ice had a measured density of
$\rho=930$ kg/m$^3$, and from (\ref{eqn:deltag2}), using $G_N=6.6742\times10^{-11}$ m$^3$s$^{-2}$kg$^{-1}$, we obtain from a
linear fit to the slope of the data points in Fig.\ref{fig:Greenland} that
$\alpha^{-1}=139\pm 5$, which equals the value of the fine structure constant  $\alpha^{-1}=137.036$ to within the errors, and
for this reason we identify the  constant $\alpha$ in (\ref{eqn:CG}) as being the fine structure constant.  Then
we arrive at the conclusion that there is indeed   `black hole' or `dark matter' dynamics within the earth,
and that from (\ref{eqn:TotalDM})  we have again for the earth that $M_{BH}/M=\alpha/2$, as is also shown in
Fig.\ref{fig:blackholes}. 

This `minimal black hole' effect must  also occur within stars, although that could only be confirmed by indirect
observations.   This effect results in  $g(r)$ becoming large at the center, unlike Newtonian gravity,  which would
affect nuclear reaction rates. This effect may already have manifested in  the solar neutrino count problem 
\cite{Davies,Bahcall}. To study this will require including the new gravity dynamics into solar models.

\begin{figure}[t]
\hspace{30mm}\includegraphics[scale=0.25]{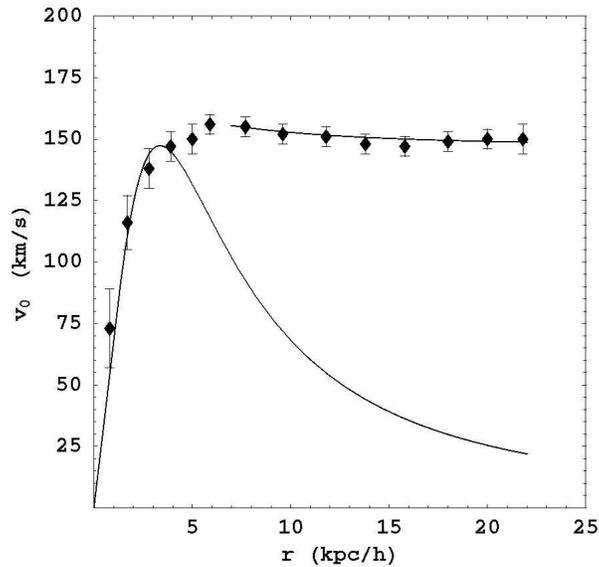}
\caption{\small {Data shows the non-Keplerian rotation-speed curve $v_O$ for the spiral galaxy NGC 3198 in km/s plotted
against radius in kpc/h. Lower curve is the rotation curve from the Newtonian theory  for an
exponential disk, which decreases asymptotically like $1/\sqrt{r}$. The upper curve shows the asymptotic form from
(\ref{eqn:vorbital}), with the decrease determined by the small value of $\alpha$.  This asymptotic form is caused by
the primordial black holes at the centres of spiral galaxies, and which play a critical role in their formation. The
spiral structure is caused by the rapid in-fall towards these primordial black holes.}
\label{fig:NGC3198}}\end{figure}

\section{\bf  Spiral Galaxies}

We now consider the situation in which matter in-falls around an existing primordial black hole. Immediately we see
some of the consequences of this time evolution: (i)  because the acceleration field falls off much slower than the
Newtonian inverse square law, as in (\ref{eqn:dmg}), this in-fall would happen very rapidly, and (ii) the resultant
in-flow would result in the matter  rotating much more rapidly than would be predicted by Newtonian gravity,  (iii) so
forming a quasar which, after the in-fall of some of the matter into the black hole has ceased, would (iv) result in a
spiral galaxy exhibiting non-Keplerian rotation of stars and gas clouds, {\it viz} the so-called `dark matter' effect.
The study of this time evolution will be far from simple. Here we  simply illustrate the effectiveness of the new
theory of gravity in explaining this `dark matter' or non-Keplerian  rotation-velocity effect.

We can  determine the star orbital speeds for highly non-spherical galaxies in the asymptotic region by
 solving (\ref{eqn:InFlowRadial}), for asymptotically where $\rho\approx 0$ the velocity field will be approximately
spherically symmetric and radial; nearer in we would match such a solution to numerically determined solutions of
(\ref{eqn:CG}).   Then (\ref{eqn:InFlowRadial})  has an exact non-perturbative two-parameter ($K$
and $R_S$) analytic solution,
\begin{equation}
v(r) = K\left(\frac{1}{r}+\frac{1}{R_S}\left(\frac{R_S}{r}  \right)^{\displaystyle{\frac{\alpha}{2}}}  \right)^{1/2}
\label{eqn:vexact}\end{equation}
This velocity field then gives using (\ref{eqn:basicg}) the  non-Newtonian   asymptotic acceleration
\begin{equation}
g(r)=\frac{K^2}{2} \left( \frac{1}{r^2}+\frac{\alpha}{2rR_S}\left(\frac{R_S}{r}\right)
^{\displaystyle{\frac{\alpha}{2}}} 
\right),
\label{eqn:gNewl}\end{equation}
applicable to the outer regions of spiral galaxies.  We then compute  circular orbital speeds  using
$v_O(r)=\sqrt{rg(r)}$  giving the   predicted  `universal rotation-speed curve'
\begin{equation}
v_O(r)=\frac{K}{2} \left( \frac{1}{r}+\frac{\alpha}{2R_S}\left(\frac{R_S}{r}\right)
^{\displaystyle{\frac{\alpha}{2}}} 
\right)^{1/2}
\label{eqn:vorbital}\end{equation}
 Because of the $\alpha$ dependent part this rotation-speed curve  falls off extremely slowly with $r$, as
is indeed observed for spiral galaxies.  This is illustrated in Fig.\ref{fig:NGC3198} for the spiral galaxy NGC 3198.

\section{\bf Interpretation and Discussion}

Sect.2 outlines a model of space  developed in \cite{Cahill, Nova} in which space has a
`substratum' structure which is in differential motion. This means  that the substratum in one
region may have movement relative to another region. The substratum is not
embedded in a deeper space; the substratum itself defines space, and requiring that, at some
level of description,  it may be approximately described by a `classical' 3-vector velocity field
${\bf v}({\bf r}, t)$. Then the dynamics of space involves specifying dynamical equations for  this
vector field. Here the coordinates ${\bf r}$ is not space itself, but a means of labelling points
in space.  Of course in dealing with this dynamics we are required to define ${\bf v}({\bf r}, t)$
relative to some set of observers, and then the dynamical equations must be such that the
vector field transforms covariantly with respect to changes of observers.  As noted here
Newtonian gravity itself may be written in terms of a vector field, as well as in terms of
the usual acceleration field ${\bf g}({\bf r}, t)$.  General Relativity also has a special class
of metric known as the Panlev\'{e}-Gullstrand metrics in which the metrics are specified by a
velocity field. Most significantly the major tests of General Relativity  involved the
Schwarzschild metric, and this metric belongs to the Panlev\'{e}-Gullstrand class. So in both
cases these putatively successful models of gravity involved, in fact, velocity fields, and so the
spacetime metric description was not essential.   As well there are in total some seven
experiments that have detected this velocity field \cite{CahillMM}, so that it is more than 
a choice of dynamical degree of freedom: indeed it is more fundamental in the sense that from it
the acceleration  field or metric may be mathematically constructed.  

Hence the evidence, both
experimental and theoretical, is that space should be described by a velocity field.  This
implies that space is a complex dynamical system which is best throught of as some kind of `flow
system'. However the implicit question  posed in this paper is that, given the physical
existence of such a velocity field, are  the Newtonian and/or General Relativity 
formalisms the appropriate descriptions of the velocity field dynamics?   The experimental
evidence herein implies that a different dynamics is required to be developed, because 
 when we generalise the velocity field modelling to include a spatial
self-interaction dynamics, the experimental evidence is that the strength of this dynamics
is determined by the fine structure constant, $\alpha$.  This is an extraordinary outcome,
implying that gravity is determined by two fundamental constants, $G$ and $\alpha$. As
$\alpha$ clearly is not in Newtonian gravity nor in General Relativity the various observational
and experimental data herein is telling us that neither of these theories of gravity is complete.
The modelling discussed here is non-relativistic, and essentially means that Newtonian gravity was
incomplete from the very beginning. This happened because the self-interaction dynamics did not
manifest in the solar system planetary orbit motions, and so neither Kepler nor later Newton were
aware of the intrinsic complexity of the phenomenon of gravity.  General Relativity was of course 
constructed to agree with Newtonian gravity in the non-relativistic limit, and so missed out on
this key non-relativistic self-interaction effect.  

Given both the experimental detection of the velocity field, including in particular the recent
discovery \cite{Nova} of an in-flow velocity component towards the sun in the 1925/26 Miller
interferometer data, and in agreement with the speed value  from (\ref{eqn:vin}) for the sun,
together with the  data from various observations herein, all showing the presence of the  $\alpha$
dependent effect, we should also discuss the physical interpretaion of the  vacuum `black hole'
solutions.  These are different in character from the so-called `black holes' of General
Relativity: we use the same name only because these new `black holes' have an event
horizon, but otherwise they are completely different. In particular the mathematical existence
of such vacuum `black holes' in General Relativity is doubtful.  In the new theory of gravity
these black holes are exact mathematical solutions of the velocity equations and
correspond to self-sustaining in-flow singularities, that is, where the in-flow speed becomes
very large within the classical description. This singularity  would then require a quantum
description to resolve and explain what actually happens there.   The in-flow does not involve any
conserved measure, and there is no notion of this in-flow connecting to wormholes etc.  The
in-flow is merely a self-destruction of space, and in
\cite{Nova} it is suggested that space is in essence an `information' system, in which case the
destruction process is easier to comprehend.   As for the in-flow into the earth, which is
completely analogous to the observed in-flow towards the sun, the in-flow singularities or `black
holes' are located at the centre of the earth, but it is unclear whether there is one such
singularity or multiple singularities.   The experimental existence of the earth-centred in-flow
singularity is indirect, as it is inferred solely by the anomalous variation of $g$ with depth,
and that this variation is determioned by the value of
$\alpha$. In the case of the globular clusters and elliptical galaxies, the in-flow singularities
are observed by means of the large accelerations of stars located near the centres of such systems
and so are more apparent, and as shown here in all case the effective mass of the in-flow
singularity is $\alpha/2$ times the total mass of these systems. It is important to note here
that even if we disregard  the theoretical velocity field theory, we would still be left with
the now well established  $\alpha/2$ observational effect.  But then this velocity field theory
gives a simple explanation for this data, although that in itself does not exclude other theories
offering a different explanation.  It is hard to imagine however how either Newtonain gravity or
General Relativity could offer such a simple explanation, seeing that neither involves $\alpha$,
and involve only $G$. As well we see that the new theory of gravity offers a very effective
explanation for the rotation charactersitics of spiral galaxies; the effect here being that the
vacuum black hole(s) at the centres of such galaxies do not generate an acceleration field that
falls off with distance according  the inverse square law, but rather according  to
(\ref{eqn:gNewl}). Remarkably this is what the spiral galaxy data shows.  This means that
the so-called `dark matter' effect is not about a new and undetected form of matter.  So the
success of the new velocity field dynamics is that one theory explains a whole range of phenomena:
this is the hallmark of any theory, namely economy
 of explanation.

\section{\bf  Conclusion}
 
The observational and experimental data confirm that the massive black holes in globular clusters and
galaxies are necessary phenomena within a theory for gravity which uses a velocity field as the fundamental degree of
freedom. This involves two constants $G$ and $\alpha$ and the data reveals that $\alpha$ is the fine structure
constant. This suggests that the spatial self-interaction dynamics, which is missing in the Newtonian theory of
gravity, may be a manifestation at the classical level of the quantum behaviour of space.  It also emerges that the
`black hole' effect and the `dark matter' effect are one phenomenon, namely the non-Newtonian acceleration caused by
 singular solutions. This effect must manifest in planets and stars, and the bore hole $g$ anomaly confirms
that for planets.  For stars it follows that the structure codes should be modified to include the new spatial self-interaction
dynamics, and to determine the effect upon neutrino count rates. The data shows that spherical systems with masses
varying over 15 orders of magnitude  exhibit  the $\alpha$-dependent dynamical effect. The non-Newtonian gravitational
acceleration of primordial black holes will cause rapid formation of  quasars and stars, explaining why recent
observations have revealed that these formed very early in  the history of the universe. In this way the new theory of
gravity makes the big bang theory compatible with these recent observations. These developments clearly have major
implications for cosmology and fundamental physics. The various experiments that
detected the velocity field  are discussed in   \cite{Nova,CahillMM}.   

This research is supported by an Australian Research Council Discovery Grant

\end{document}